\documentclass{ws-procs975x65}
\def\beq{\begin{equation}}
\def\eeq{\end{equation}}
\begin{document}

\title{Ultra-stripped supernovae and double neutron star systems}
\author{Thomas M. Tauris}

\address{AIfA, University of Bonn\\
53121 Bonn, Germany\\
E-mail: tauris@astro.uni-bonn.de}

\begin{abstract}
The evolution of close-orbit progenitor binaries of double neutron star (DNS) systems leads to supernova (SN) explosions of ultra-stripped stars. 
The amount of SN ejecta mass is very limited from such, more or less, naked metal cores with envelope masses of only $0.01-0.2\;M_{\odot}$. 
The combination of little SN ejecta mass and the associated possibility of small NS kicks is quite important for the characteristics of the 
resulting DNS systems left behind. Here, we discuss theoretical predictions for DNS systems, based on Case~BB Roche-lobe overflow prior to ultra-stripped SNe, 
and briefly compare with observations. 
\end{abstract}

\keywords{supernovae; neutron stars; stellar evolution; binaries; pulsars}

\bodymatter

%%%%%%%%%%%%%%%%% now a standard article style for the most part

\section{Case~BB RLO from Evolved Helium Stars in Close Binaries}
Prior to the formation of the second neutron star (NS) in a double neutron star (DNS) system \cite{tv06}, the progenitor of the exploding star --- 
a helium star (naked core) emerging from a common envelope (CE) \cite{pac76,ijc+13} evolution ---  is expanding during helium shell burning.
For close binaries, this will cause a last and significant phase of mass transfer, so-called Case~BB Roche-lobe overflow (RLO) \cite{hab86a,dpsv02,dp03,ibk+03},
which is important for two reasons: i) it strips the outer layers of the helium star very efficiently, leaving an almost
naked metal core prior to the supernova (SN) explosion \cite{tlm+13,tlp15}, and ii) it causes the first-born NS in the binary to
accrete material and become a recycled radio pulsar \cite{dpp05,tlp15}. 
A similar episode of Case~BB RLO from post-CE binaries can also produce a recycled NS in a binary
with a massive white dwarf (WD) companion \cite{tlk12,ltk+14}. 

Some of the remaining questions to investigate in relation to this recycling process are: 
 the decay of the B-field of the accreting NS, 
 its accretion efficiency,
 and how the excess material (transferred towards the NS at a highly super-Eddington rate
 from the donor star) is ejected from the vicinity of the NS.

\section{NS Accretion prior to and after Case~BB RLO}
When discussing the recycling of the first-born NS in a DNS system, we must first identify the possible phases
of accretion onto this star. The following five phase are relevant \cite{tau+16}: 
  I) wind accretion in the high-mass X-ray binary (HMXB) phase, 
 II) accretion during the CE phase, 
III) wind accretion from the naked helium (Wolf-Rayet) star, 
 IV) Case~BB RLO from the evolved helium star, and 
  V) accretion of SN ejecta from the explosion of a very nearby naked core forming the second NS.

The amount of wind accretion in phases~I and III is constrained by the lifetime of the massive hydrogen-rich star ($\sim\!10\;{\rm Myr}$) 
and the subsequent naked helium star evolution ($<1\sim2\;{\rm Myr}$), respectively, their wind mass-loss rates and the efficiency of NS accretion. 
For fast (supersonic) winds, the Bondi-Hoyle-like accretion is rather inefficient. The accretion is
more efficient for (sub)giant donors with slower winds. In some HMXB sources, so-called beginning atmospheric
RLO might be at work. However, this later stage of evolution is relatively short lived \cite{sav78}. Furthermore,
the accretion efficiency may be low as a consequence of propeller effects \cite{is75}.
These circumstances are also evident from X-ray luminosity functions of Galactic HMXBs \cite{ggs02}, which indicate that less than about 10\% of
all HMXBs have a luminosity $>10^{37}\;{\rm erg\,s}^{-1}$, corresponding to an accretion rate of
$\sim\!10^{-9}\;M_{\odot}\,{\rm yr}^{-1}$. 
 All taken together, we estimate that the total amount of material accreted by the NS in these two phases
is less than $0.01\;M_{\odot}$. 

The accretion onto a NS embedded in a CE (phase~II)  has to be quite limited as evidenced from measurements of radio
pulsar masses in post-CE binaries. A possible reason for this has recently been demonstrated \cite{mr15b} to be caused by
the density gradient across the gravitational capture radius imposing a net angular momentum to the flow, thereby
significantly decreasing the accretion efficiency compared to Bondi-Hoyle accretion. It is likely that the 
amount of accreted matter during a CE phase could be as low as $0.01\;M_{\odot}$. 
Given that a stable accretion disk may not form around the NS within a CE, it is uncertain whether this small amount
of matter will contribute to any recycling (spin-up) of the accreting NS at all.
  
Accretion of ejecta matter from the SN of a companion star (phase~V) is expected to be highly insignificant in general, except for
a few rare and very finetuned cases \cite{frr14}, and we will disregard this possibility in further discussions.

\section{NS Recycling via Case~BB RLO}
Case~BB RLO has recently been demonstrated \cite{tlp15} to result in accretion onto the NS of $5\times10^{-5}-3\times10^{-3}\;M_{\odot}$
for Eddington-limited accretion, and up to $10^{-2}\;M_{\odot}$ if allowing for accretion above the Eddington
limit by a factor or three. The gain in spin angular momentum of the accreting NS can be expressed as \cite{tlk12}:
\begin{equation}
  \Delta J_\star = \int n\,(\omega,t)\,\dot{M}(t)\,\sqrt{GM(t)\,r_{\rm mag}(t)}\,\xi(t)\,dt,
\end{equation}
where $n$ is the dimensionless accretion torque, $\omega$ is the fastness parameter, $t$ is time, $\dot{M}$
is the NS mass accretion rate, $G$ is the constant of gravity, $M(t)$ is the mass of the NS, $r_{\rm mag}$
is the radius of the magnetosphere, and $\xi$ is a parameter depending on the geometric flow of the 
material ($\approx 1$ for accretion from a disk).
This integral can roughly be approximated as a simple correlation between the equilibrium spin period in ms, $P_{\rm ms}$ and the 
(minimum) amount of mass accreted, $\Delta M_{\rm eq}$:
\begin{equation}
  P_{\rm ms} = \frac{(M/M_{\odot})^{1/4}}{(\Delta M_{\rm eq}/0.22\;M_{\odot})^{3/4}}
\end{equation}
Hence, we find that the abovementioned range of mass accreted via Case~BB RLO corresponds to spinning up the NS to a 
resulting spin period of $11-350\;{\rm ms}$ \cite{tlp15}.
For a comparison, the observed range of spin periods of recycled NSs in DNS systems is currently spanning between  $23-185\;{\rm ms}$.

\section{Correlation between DNS Orbital Period and Spin Rate}
\begin{figure}[t]
\begin{center}
\includegraphics[width=12.0cm, angle=0]{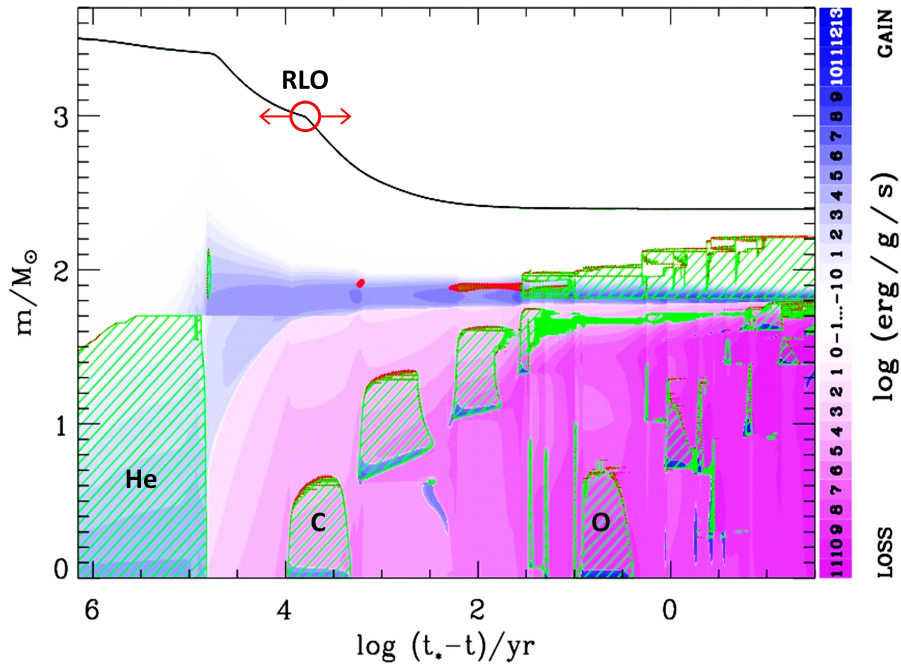}
\end{center}
\caption{Kippenhahn diagram of a $3.5\;M_{\odot}$ helium star undergoing Case~BB RLO, showing 
the evolving cross section of the star, in mass coordinate on y-axis, as a function of remaining
calculated lifetime on the x-axis. The green hatched areas denote zones with convection and the
intensity of the blue/purple colour indicates the net energy-production rate.
The total mass of the star is shown by the solid black line. Depending on the initial orbital period,
the onset of the RLO (marked by a red circle) will occur either at an early or late evolutionary state of the helium star.
This figure was adapted from Ref.~[5].}
\label{fig:Kippenhahn}
\end{figure}

\begin{figure}[t]
\begin{center}
%\vspace{-1.0cm}
\includegraphics[width=9.0cm, angle=-90]{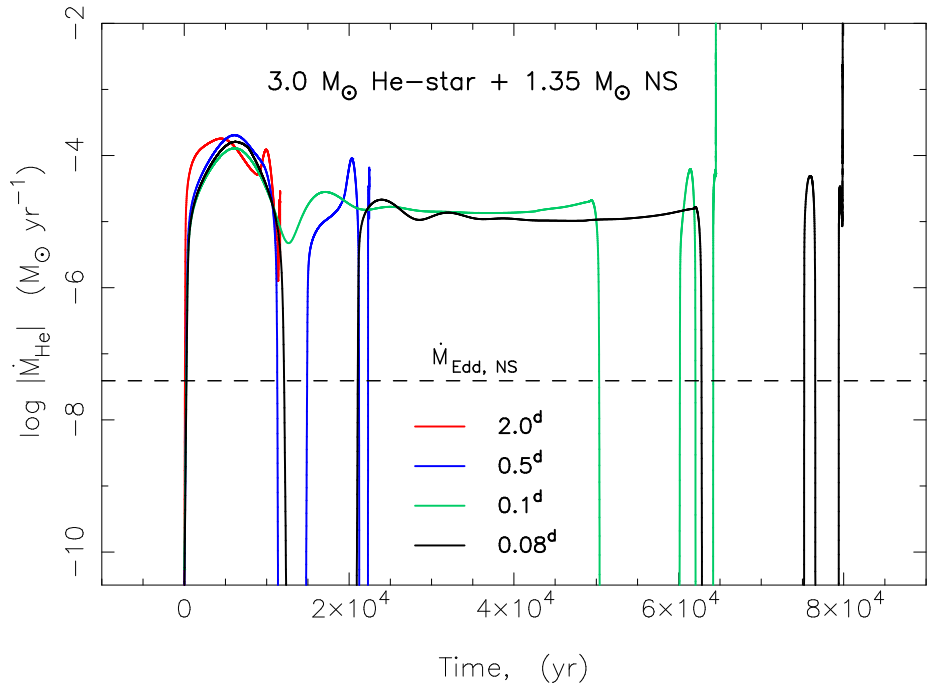}
\end{center}
\caption{Mass-transfer rate as a function of time for $3.0\;M_{\odot}$ helium star donors with different values of
initial orbital period at the onset of Case~BB RLO. The time $t=0$ is defined at the beginning of RLO for all systems.
The duration of the RLO (and thus the amount of material transferred onto the NS) is seen
to decrease with increasing orbital period. Thus recycling is inefficient in wider~orbits, leading to slower
spin periods of recycled pulsars in wide-orbit DNS~systems. 
This~figure was adopted from~Ref.~[5].}
\label{fig:Mdot_age}
\end{figure}

Given the relation between amount of accreted material and equilibrium spin rate of the recycled pulsar, we can predict \cite{tlp15,tau+16} 
a correlation between the orbital period and the spin period of recycled pulsars in DNS systems, since the amount of 
accreted mass is strongly dependent on the orbital period of the binary at the onset of Case~BB RLO.
This is illustrated in Fig.~\ref{fig:Kippenhahn}.
Detailed calculations of Case~BB RLO showing how the duration of the mass-transfer phase decreases with increasing
initial orbital period is shown in Fig.~\ref{fig:Mdot_age}. 
The wider the initial orbit of the NS--helium star system, the more evolved is the helium star when it fills its Roche lobe 
and the shorter is its remaining lifetime before it collapses and produces a SN explosion. 
Therefore, in wide binaries little mass is transfered to the NS prior to core collapse, and thus the recycling process
is relatively ineffective. The spin period of such a {\it marginally recycled pulsar} will remain large. 
PSR~J1930$-$1852 \cite{srm+15} is an example of such a system. It has an orbital period of 45~days and a slow spin period of 185~ms.
\begin{figure}[t]
\begin{center}
%\vspace{-1.0cm}
\includegraphics[width=8.0cm, angle=-90]{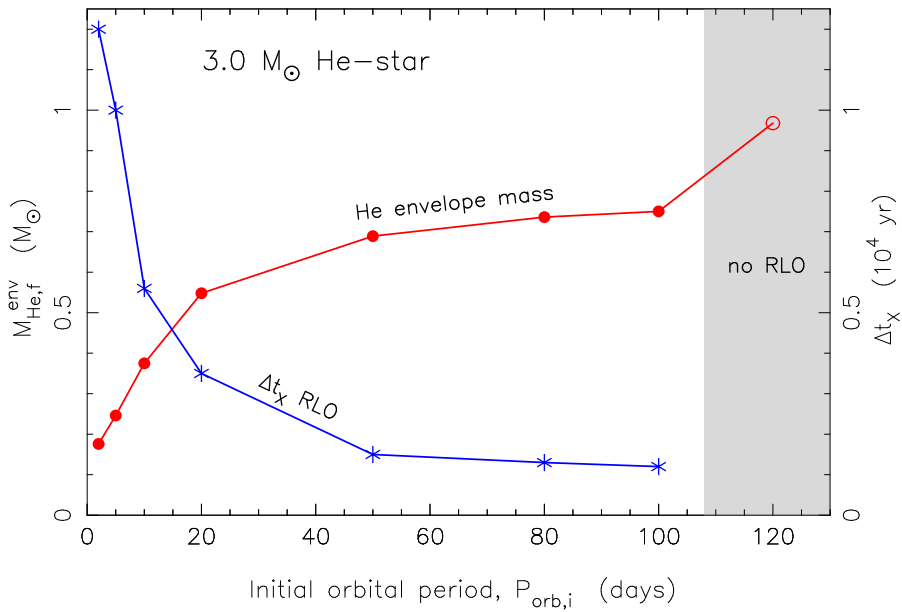}
\end{center}
\caption{Duration of Case~BB RLO (decreasing blue curve), and remaining helium envelope mass (increasing red curve), prior to the iron core-collapse SN
as a function of initial orbital period. Wide-orbit (post-SN) DNS systems are thus expected to host 
pulsars which are only marginally recycled (as a result of their short-duration RLO) and which, in average, have higher eccentricities 
(as a result of their larger envelope mass ejected in the SN) than pulsars in close-orbit systems.
This figure was adopted from Ref.~[5].}
\label{fig:wideDNS}
\end{figure}
In Fig.~\ref{fig:wideDNS} we have plotted the duration of the Case~BB RLO prior to the SN explosion, as well
as the remaining amount of helium envelope mass at the moment of the SN, as a function of initial orbital period at 
the onset of Case~BB RLO. Since wide-orbit pre-SN systems are also expected to produce wide-orbit DNS systems after the SN explosion 
(especially if NS kicks are small, see discussion below) we expect that these systems, which are clearly seen to experience
short-duration RLO and thus limited recycling, will produce relatively slow spinning recycled pulsars. Similarly, we expect the
fastest spinning recycled pulsars in DNS systems to be found in close-orbit systems, such as PSR~J0737$-$3039 \cite{bdp+03} which has
a spin period of 23~ms and an orbital period of 2.4~hr.

\section{Ultra-stripped SNe, NS kicks and DNS eccentricities}
Ultra-stripped SNe can be either electron-capture SNe (EC~SNe) or iron core-collapse SNe (Fe~SNe). 
It has been shown that EC~SNe are expected to have small explosion energies \cite{dbo+06,kjh06} and that they revive the stalled SN shock
on a short timescale compared to the timescales of non-radial hydrodynamic instabilities producing strong anisotropies \cite{plp+04,jan12}. 
Their resulting kick velocities are most likely $<50\;{\rm km\,s}^{-1}$, i.e. significantly smaller than the average
kick velocities of the order $400-500\;{\rm km\,s}^{-1}$ imparted on young pulsars \cite{hllk05}.
It has also been argued \cite{tlp15} that ultra-stripped Fe~CCSNe might (in general, but not always) produce small NS kicks.
The reasons for this are: i) their small amount of ejecta mass, compared to standard SNe, which leads to a weaker gravitational
tug on the proto-NS \cite{jan12}, and ii) the small binding energies of their envelopes allowing for fast ejection, potentially
before large hydrodynamical anisotropies can build up.
Recent modelling of ultra-stripped SN explosions of evolved CO-stars \cite{suw15} confirms that their kick velocities are expected to be small in general. 
Nevertheless, in a few cases (PSR~B1534+12 and PSR~B1913+16) kicks of the order $300\;{\rm km\,s}^{-1}$ seem necessary to explain
the orbital parameters of the DNS systems. Hence, we conclude that whereas ultra-stripped SNe might produce small
kicks in general, larger kicks are possible in more rare cases -- possibly caused by explosions of relatively more massive iron cores. 

Finally, it is expected that ultra-stripped SNe (at least those which produce small kicks) result in relatively small DNS eccentricities in general.
The reason for this is their small ejecta masses. It should be noted, however, that even small kicks $(50\;{\rm km\,s}^{-1})$ can in some cases
result in large eccentricities, depending on the direction of the kick. In average, we do expect \cite{tlp15} a trend of increasing
eccentricities with increasing orbital period of DNS systems, given that the exploding star in wider pre-SN systems will possess (and thus eject)
a larger envelope mass, cf. Fig.~\ref{fig:wideDNS}.

Detailed modelling of the recycling and spin-up of the accreting NS during Case~BB RLO has to be investigated further.
Initial studies \cite{tlk12,tlp15} indicate that the computed spin periods of the recycled NSs in DNS  systems are indeed 
in good agreement with the observed values of $23-185\;{\rm ms}$. Many (but not all) \cite{tau+16} of the DNS systems also have small
eccentricities and small systemic velocities (reflecting small kicks). Future modelling of the explosions are needed to investigate this further.

\end{document}